# EyeSec: A Retrofittable Augmented Reality Tool for Troubleshooting Wireless Sensor Networks in the Field

Martin Striegel, Carsten Rolfes, Johann Heyszl
Fraunhofer AISEC
Garching near Munich, Germany

{martin.striegel, carsten.rolfes, johann.heyszl}@aisec.fraunhofer.de

Fabian Helfert, Maximilian Hornung, Georg Sigl
Technical University of Munich
Munich, Germany

{fabian.helfert, maximilian.hornung, sigl}@tum.de

**Abstract**

Wireless Sensor Networks (WSNs) often lack interfaces for remote debugging. Thus, fault diagnosis and troubleshooting are conducted at the deployment site. Currently, WSN operators lack dedicated tools that aid them in this process. Therefore, we introduce *EyeSec*, a tool for WSN monitoring and maintenance in the field. An Augmented Reality Device (AR Device) identifies sensor nodes using optical markers. Portable *Sniffer Units* capture network traffic and extract information. With those data, the AR Device network topology and data flows between sensor nodes are visualized. Unlike previous tools, EyeSec is fully portable, independent of any given infrastructure and does not require dedicated and expensive AR hardware. Using passive inspection only, it can be retrofitted to already deployed WSNs. We implemented a proof of concept on low-cost embedded hardware and commodity smart phones and demonstrate the usage of EyeSec within a WSN test bed using the 6LoWPAN transmission protocol.

## 1 Introduction

Advances in low-power sensing, embedded computing and wireless protocols drive the dissemination of wireless sensor networks (WSNs). After deployment, sensor nodes collect sensor data and transmit those to a server for processing and storage. The server is often connected to an existing IT infrastructure using a gateway.

Similar to enterprise-class IT devices, the WSN is usually monitored by capturing network traffic at the gateway and displaying node and link status to the network operator at a central control terminal [12, 29]. However, after a device malfunction has been detected, situations are different. While enterprise-class IT devices can be managed actively using e.g. *Simple Network Management Protocol (SNMP)*, sensor node behavior can only be observed *passively*. Sensor nodes typically lack network management protocols, because flash memory is scarce, firmware must be sleek and additional network traffic, which could interfere with WSN operations, must be avoided. Hence, if sensor node failure is reported by the central monitoring system, an operator is sent into the field to pinpoint and fix the problem on-site. To be able to provide the operator with information on the network at the location of deployment, Turon et al. and Bokde et al. describe hand-held devices [30, 11]. Those devices obtain information from the gateway. This is disadvantageous, as the hand-held devices require a permanent connection to the gateway and their operativeness depends on those of the gateway. To become independent of the gateway as network traffic source, separate and passive capture devices have been proposed [23, 22, 32]. Consisting of mobile capture or *sniffer* nodes, they can be deployed temporarily to overhear all WSN traffic.

Combining mobile capture networks and hand-held devices for visualization permits the operator to work independently of existing infrastructure. However, by just shifting the visualization of network data from a central terminal to a hand-held device does not tackle specific problems encountered by the operator. While a complete view on the WSN such as text-based traffic statistics and large network graphs can be displayed at a hand-held device, the operator needs only limited information targeted at solving a specific problem. Instead, the operator needs to map *digital* device representation, i.e. network addresses, and the *visual* device representation of the physical sensor node in front of him. This permits to identify the physical device causing problems in the network. In turn, while manipulating a physical device, the operator needs to see the effects on the digital world, e.g. if a network connection can be restored.

Currently, there exists no tool designed to provide the operator in the field with just the information needed to debug wireless sensor networks. To fill this gap, we propose *EyeSec*, a tool utilizing *Augmented Reality (AR)* and tailored towards the operator's needs. EyeSec includes Augmented Reality Devices (*AR Devices*), which detect and identify physical sensor nodes using Quick Response (QR) code markers. Portable and extensible *Sniffer Units* capture network traffic and extract digital information. Data from the visual and digital domain are merged and stored at a portable *Backend*. An AR Device obtains consolidated information from the Backend and superimposes physical sensor nodes with this information. Data flows and connectivity between sensor nodes are visualized. Unlike centralized network visualization solutions, which display the full network, we exploit the operator's physical proximity to sensor nodes to limit displayed information to exactly those he is interested in. EyeSec is designed such that Sniffer Units and Backend can be installed ad-hoc at a WSN deployment site and removed after troubleshooting is finished, neither requiring changes to sensor node firmware nor interfering with WSN operations at any time.

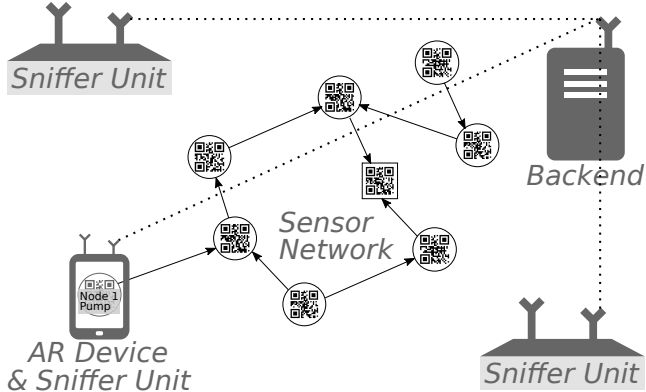

**Figure 1. Using EyeSec to monitor a WSN.**

Our main contributions are:

- The first completely mobile WSN monitoring system, which permits the operator to work independently of any existing infrastructure, called EyeSec.
- Passive operation, i.e. no modifications to sensor node or gateway firmware need to be made. This permits EyeSec to be retrofitted to already deployed WSNs without introducing additional sources of error.
- Modular and protocol-agnostic design, so extra transmission protocols can be added easily. EyeSec utilizes off-the-shelf radio transceivers and Android OS smart phones, which are available widely and cheaply.

The remainder of this paper is structured as follows: In Section 2, we detail requirements and design considerations for building a retrofittable and protocol-agnostic AR system. In Section 3, we show how the design has been implemented in hardware and software. In Section 4, we evaluate EyeSec in a real sensor network using 6LoWPAN transmission protocol. In Section 5, we compare EyeSec to similar tools. Section 6 concludes the paper and outlines future work.

## 2  Design of EyeSec

In this section, the design of EyeSec is presented. Firstly, we give an overview of the hardware units, which are part of EyeSec, and how they communicate in Section 2.1. In Section 2.2, we outline how EyeSec is used and how usability is incorporated into the design. Each of the hardware units, Sniffer Unit, Backend and AR Device, is assigned a processing pipeline. We describe those processing pipelines in detail in Sections 2.3 to 2.6. Finally, in Section 2.7 we discuss security design.

### 2.1  System Design and Communications

EyeSec consists of three hardware units called Sniffer Unit, Backend and AR Device. They have been designed modularly, such that hardware units can be merged or split depending on the size of the WSN to be monitored. Any combination of Sniffer Unit, Backend and AR Device can be utilized. Figure 1 shows those hardware units applied to a WSN. The WSN is displayed in the center. Every sensor node has an optical marker. Two dedicated Sniffer Units, shown to the top left and bottom right, have been deployed among sensor nodes and passively capture traffic from the sensor network. They extract data from traffic and transmit those data to a Backend, where information is stored. The AR Device is a hand-held device carried by the operator. It reads the optical marker of a sensor node, from which it can derive the identity of the sensor node. In addition, it has an integrated Sniffer Unit. The AR Device fetches data from the Backend and superimposes a sensor node with those data, e.g. the name and location of the node.

Hardware units communicate using Wi-Fi. The Backend creates a protected Wi-Fi network, which Sniffer Units and AR Devices join. We chose Wi-Fi to connect EyeSec devices, as Wi-Fi data rates exceed those of low-power transmission protocols used in WSNs. Thus it is ensured, that even in large-scale WSNs with high data transmission rates all extracted information can be exchanged in time. This ensures that communications do not become a bottleneck. Additionally, since Wi-Fi can be operated in either 2.4 GHz or 5 GHz band, we can choose a band which does not interfere with WSN operations. This ensures, that EyeSec operates truly passively.

As all devices are part of the same network, they can be time-synchronized using *Network Time Protocol (NTP)*. The Backend is configured as NTP server and can be equipped with a real-time clock module. Sniffer Units and AR Devices are NTP clients, which fetch time information from the server. Hence, we can utilize time stamps in data acquisition.

### 2.2  Usage and Usability Design

The modular design of EyeSec permits the operator to split and combine hardware units as needed to maximize usability. In the first exemplary use case, the operator is sent at the deployment site of a small WSN to pinpoint and troubleshoot an error.

For this task, he can utilize an AR Device with an integrated Sniffer Unit, as pictured to the bottom left in Figure 1. Even the Backend can be merged into the combined Sniffer Unit/AR Device, omitting the need to place any separate hardware unit and providing the network operator with a single hand-held tool. While he might not be able to capture all network traffic with the single AR Device/Sniffer Tool, the operator is still able to inspect the sensor node of interest and its neighborhood. The single combined Sniffer Unit/AR Device is sufficient to capture traffic to and from the particular sensor node, which is currently inspected with the AR Device. Using this setup, the operator benefits from high mobility and zero setup time.

In another exemplary use case, a new WSN has been deployed. The operator is sent on-site to confirm that all sensor nodes work as expected. As this inspection likely has to be repeated several times, until it is verified that long-time stability of the network is given, the operator places multiple dedicated Sniffer Units such that they can capture all WSN traffic. Additionally, he installs a separate Backend. Depending on the estimated duration of monitoring, those devices can be powered by battery or grid power. Utilizing the hand-held AR device, the operator can approach sensor nodes and check their status and connectivity. After having confirmed long-time stability, he can remove the Sniffer Units and re-use them for observing another WSN. At any future point in time, after having conducted modifications or upon encountering failure behavior in a WSN, EyeSec can be re-applied for monitoring and troubleshooting. Again, the portable and modular design of EyeSec is beneficial as the operator can integrate it easily into his work flow rather than being forced into a certain work flow imposed by restrictions of the tool.

### 2.3  Digital Device Representation Pipeline

Each Sniffer Unit processes a *Digital representation pipeline*, shown to the left hand side in Figure 2. Its input stage captures network traffic. Next, it extracts communicating parties and associated information from captured traffic. Extracted data are then sent to the information storage at the Backend.

Block *D1 Passive network traffic capture* in Figure 2 needs to account for the diversity of transmission protocols, which can be encountered in WSNs. Among those are e.g. *Bluetooth (Mesh)*,

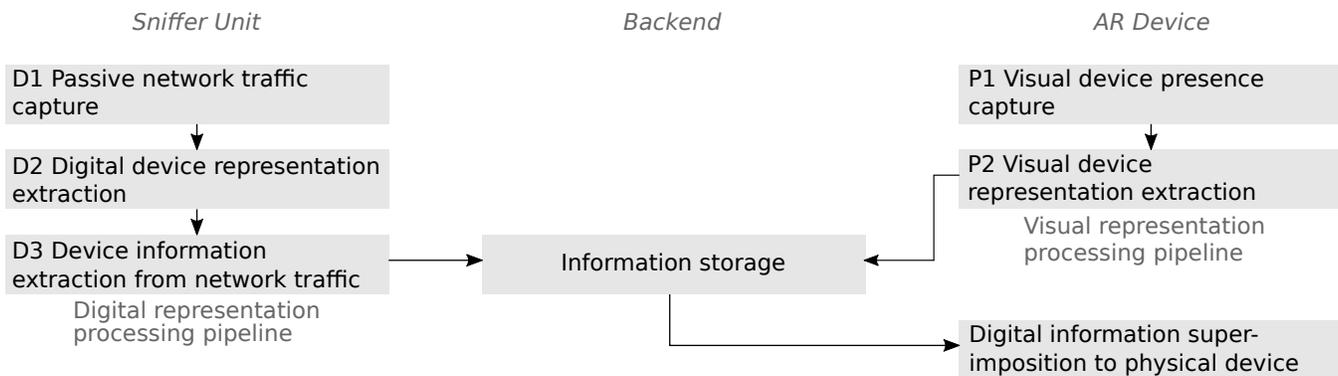

**Figure 2. EyeSec processing pipelines mapped to Sniffer Unit, Backend and AR Device.**

*802.15.4*-based protocols such as *ZigBee 3.0* and *6LoWPAN*, *Wi-Fi* or *LoRaWAN*.

Since OSI Layers 1 and 2 (Physical and Data Link Layer) differ among transmission protocols, we need hardware capable of passively capturing the transmission protocol of interest. Here, either a Software-Defined Radio (SDR) or dedicated transceivers can be used. While the former offer great flexibility, the latter are more cost efficient, smaller and have lower energy consumption. Thus for our mobile application, they are preferable over SDRs. Most capture hardware outputs packets in *PCAP* format, which is the de-facto standard for packet processing. This permits using established packet handling tools such as *Wireshark* [9].

In block *D2 Digital device representation extraction*, the *digital representation* of a device is obtained from captured network data. This representation consists of its identifier in the network and associated digital information. Again, we have to deal with various transmission protocols used in WSNs which use heterogeneous primary addressing. For example, Bluetooth uses hardware addresses (Media-Access-Control addresses (MAC)) to deliver packets. LoRaWAN uses a Device-EUI (Extended Unique Identifier), often but not necessarily set to the MAC address. In contrary, 6LoWPAN utilizes IPv6 addressing. To homogenize this building block, we need a common digital device identifier, which must be derived from network traffic. This must be accomplished using only passive inspection of network traffic, since EyeSec shall neither require changes to sensor node firmware, nor interfere with WSN operations.

The hardware address (MAC) of a device is a suitable choice, as it is present in every networked device and typically static and unique within a network. Deriving the MAC addresses of network members from the primary addressing depends on the actual transmission protocol. Hence, processing block *D2* must be adapted to the network protocol of interest.

In block *D3 Device information extraction from network traffic*, we obtain information such as packets sent and received. In mesh networks, direct neighbors of a sensor node as well as the hops a packet takes are of interest. We treat hops as individual packets between neighboring nodes identified by MAC addresses, which we have derived from the primary network addressing scheme in block *D2*. In WSNs, which do not use mesh routing, EyeSec considers the hop-size to be one. Since the MAC address is known already from block *D2*, we can directly assign network information to a particular sensor node.

EyeSec has been designed to utilize multiple Sniffer Units to fully cover the sensor network. However, their receive radii might overlap. Thus, a single packet might be captured by multiple Sniffer Units. If no actions were taken, the Backend, to which information extracted from packets are sent, would be cluttered with duplicate information, falsifying traffic statistics as well as information displayed to the operator. To prevent duplicates in the database, we use the following procedure. Upon packet capture, the Sniffer Unit calculates a hash value of all data below the Network Layer using the Message-Digest Algorithm 5 (MD5). Those lower packet fields are static during the journey of a packet through the network. The Sniffer Unit transmits the hash value together with the packet data and the capture time stamp to the Backend. Using time stamps is possible since the Sniffer Units and the Backend are time-synchronized over the common wireless network, which they are all part of. Upon receiving a packet from a Sniffer Unit, the Backend can search its database using the MD5 hash value as an index, which permits efficient search. If a packet indexed with the received hash is found in the database, time stamps are compared. If the difference of time stamps is smaller than $\varepsilon$, the packets are considered duplicates and the freshly received packet is discarded. Typically, $\varepsilon$ is chosen in the range of milliseconds. This accounts for varying packet travel times between message source and the Sniffer Units and especially for the precision limits of the time synchronization protocol NTP. Else, if the difference of time stamps is larger than $\varepsilon$, the received packet is added to the database.

We deliberately chose to let the Sniffer Unit create the MD5 hash and always send it together with the data to the Backend. One might argue, that those computations are wasted in case the Backend considers data as duplicate and discards it. While this is true, this design choice shifts computational burden from the Backend to the Sniffer Unit, preventing performance bottlenecks at the Backend. If packets were transmitted without the pre-computed MD5 hash, the Backend would be tasked with calculating hash values. With an increasing amount of Sniffer Units, the Backend would face high computational burden, which would ultimately require faster hardware, increase power consumption and reduce its portability. Thus, having the Sniffer Units calculate the hash value offers scalability, as this permits the operator to use as many Sniffer Units as needed to capture all traffic in the WSN.

EyeSec is protocol-agnostic in the sense that it is neither dependent on special message types nor information only available in a particular transmission protocol. With minor modifications to the digital device representation, extra transmission protocols can be added easily. Its homogenized output permits, that processing blocks, which are assigned to the Backend and the AR Device, need not be adjusted to a particular transmission protocol.

## 2.4 Visual Device Representation Pipeline

Every AR Device executes a *Visual representation processing pipeline*, which is shown on the right hand side in Figure 2. In step *P1 Visual device presence capture*, the actual presence of a device of interest needs to be detected. In the second step *P2 Visual device*

*representation extraction*, the detected device needs to be identified. This distinction between presence detection and identification has an impact on the technologies used to perform those steps. Using for example *image recognition* with mature frameworks such as *OpenCV*, a device of interest can be detected reliably without needing optical markers [3]. However, identification of devices is difficult since networked devices share a similar visual appearance and are not distinguishable from another [15]. Additionally, image processing requires expensive computations, quickly draining the battery of the AR Device. Hence, *marker-based* approaches are favorable over markerless solutions. Quick Response (QR) codes are optical markers, which permit both detection and identification of devices. They can be detected and read reliably by a camera. This offers great flexibility in choosing the capture hardware. Each sensor node needs to be supplied with a QR code. In the QR code, the MAC address of the device is embedded. Upon identifying a new sensor node by its QR code, the newly found device is announced to the Backend. There, we now have the MAC address as common device representation for both the digital and the visual world.

We are aware of the administrative overhead introduced by the addition of QR codes. However, most of the time, industrial devices are supplied with a printed sticker containing information on the device, so the QR code can be added easily. Additionally, unlike other approaches such as a sensor node blinking its MAC address in a Morse code way using an LED, as discussed in [27], we do not need modifications to the sensor node firmware. This brings high acceptance, as extending the firmware always comes at the risk of introducing new errors. Thus, using QR codes enables retrofitting of EyeSec to already deployed WSNs.

## 2.5 Information Storage at Backend

The Backend is responsible for *Information storage*, shown in the center of Figure 2. It receives extracted device information, such as message transfers from one or more Sniffer Units and visual device information from AR Devices.

The storage of the Backend is updated whenever new devices have been identified in steps *D2* and *P2*. The storage shall be the single central instance in the system, where information is merged and supplied. It must be ensured that no duplicate information is stored. Thus, the design of this storage is crucial for the overall system performance.

As visual and digital device representation are linked by the devices MAC address, it is used as identifier for consolidating data from both domains. A suitable choice for storing information at the Backend is a database, which features fast write and read operations and is capable of handling simultaneous accesses, for example write operations of multiple Sniffer Units. Duplicate information can occur, if e.g. the same QR code is scanned twice or if a single network packet is captured by more than one Sniffer Unit. In the case of scanned QR codes, the Backend searches its database whether the MAC address derived from the QR code is already present. To prevent duplicate packets being stored, Sniffer Units send their data to the storage combined with an MD5 hash identifier and a time stamp. The procedure to check, whether a packet received from a Sniffer Unit is a duplicate, has been described in section 2.3.

## 2.6 Output at AR Device

*Digital information superimposition to physical device* is the terminating processing block. It is shown on the bottom right in Figure 2. In this step, the AR Device obtains information from the Backend and annotates sensor nodes with those.

EyeSec shall support common hardware for AR Devices to provide flexibility in the implementation and remove the need of acquiring expensive dedicated hardware. An AR Device needs to be able to acquire combined visual-digital data from the information storage without requiring wired connections. Further, it must be able to detect and extract information from QR codes placed on sensor nodes. Simultaneous detection of multiple QR codes is needed, as this is a common situation encountered whenever multiple sensor nodes are within the camera image. Lastly, it needs to annotate a device identified by a QR code with the visual-digital data, using e.g. an overlay.

Due to their ubiquity and compliant hardware, smart phones are suitable AR Devices. However, one could also use a tablet computer or a notebook equipped with a web cam. We decided against using AR Headsets, as they are currently more expensive and less common than the solutions mentioned before. To create the annotation overlay on sensor nodes, EyeSec uses custom line draws. We could have also used established solutions such as *Vuforia*, *EasyAR* or *ARCore*. However, the EyeSec app only uses basic features of AR (recognition of QR codes and overlay view rendering). Thus, most features of the advanced AR SDKs are simply not necessary and would just clutter the app. Additionally, we are offered great flexibility and neither need expensive licensing nor cloud access, as it would be the case with the solutions mentioned beforehand.

## 2.7 Security Design

Besides 'natural' failure due to sensor node death by e.g. drained battery or bugs in the firmware, WSN operability can also be impaired by a cyber attack. For example, mesh networks can be subject to routing attacks, which cause network traffic to be misdirected or dropped [19, 31, 13]. Besides attacking routing protocols, WSN integrity can be assaulted by spoofing, i.e. copying a sensor nodes identity to a malicious sensor node. This leads to network traffic being attributed or directed falsely by the routing protocol.

While the MAC address is a convenient choice for sensor node network identity, it can be spoofed easily by an attacker, as it is publicly known. To prevent identity spoofing, *cryptographically secure identities* for sensor nodes need to be used. Such identities can be built using e.g. secret keys, public-private key pairs managed by a public key infrastructure (PKI) or physically uncloneable functions (PUFs). If a secure identity is available, messages can be encrypted and then authenticated. The legitimate message sink node verifies that the sender really is the entity it claims to be. If this verification succeeds, the receiver decrypts the message contents. Else, it discards the message.

EyeSec has been designed such, that it can be used with secured WSNs. To be able to verify the secure identity of a device, a Sniffer Unit needs to be supplied with this information. For example, each sensor node could have a public-private key pair. The private key is used to sign messages, while the public key is used to validate signatures by message recipients. Thus, EyeSec needs to know the public key of every sensor node, e.g. by giving EyeSec access to the Certificate Authority (CA), which has confirmed that a specific public key belongs to the address of a particular sensor node. Using the public key of a sensor node, EyeSec is able to validate message signatures of captured traffic and thus confirm the identity of a sensor node. If validating the authenticity of a sensor node fails, EyeSec concludes that the node uses a spoofed network address and issue a warning to the operator, revealing spoofing attacks.

Targeting EyeSec operations, an attacker can attempt to copy or forge the sensor nodes visual representation, i.e. the QR codes, together with its network representation, the MAC address. EyeSec creates a single visual device representation when the operator scans the first QR code. Further, it sniffs traffic from both the legitimate and the malicious sensor node. As EyeSec has been supplied with public keys, it validates the signatures of all messages. Assuming the attacker has not stolen the secret key of the sensor node, whose identity has been copied, the malicious node is not able to forge message signatures. EyeSec tries to apply the public key of a legitimate sensor node to validate the signature but will fail, as the public key used does not match the secret key used by the malicious node to sign messages. As a result, EyeSec reports the failed signature validation to the operator.

In the case that the malicious sensor node does not transmit messages, the signature validation is of no use. Traffic originating from the sensor node, whose digital and visual representation has been copied, pass the signature validation as public and private key match. However, if no countermeasures were applied, EyeSec would visualize traffic originating from the legitimate sensor node at the copied node, too. To prevent this, EyeSec issues a warning upon detecting duplicate QR codes. In order to distinguish the legitimate from the malicious sensor node, signal strength measurements can be conducted. Approaching the legitimate sensor node, the signal strength of sniffed packets attributed to this sensor node increases. Moving from the legitimate to the malicious node, signal strength of those packets drops since the legitimate node either sends no packets at all or packets originating from it fail the signature validation and are thus discarded by EyeSec.

In the case that a QR code is copied, but the malicious sensor node uses a network representation different from any other encountered in the network, the situation is slightly different. The malicious node can now send messages using its own network representation. However, those messages fail the signature check, as the public key matching this network address is not known to EyeSec. Network traffic from the legitimate sensor node would still be visualized at the malicious node, but as in the example above, EyeSec detects duplicate QR codes and issues a warning.

If an attacker uses a forged QR code, but has copied the network address of a legitimate sensor node, the operator can add the malicious nodes visual representation to the Backend. Traffic originating from the malicious node fails the signature check and EyeSec issues a warning to the operator. As the visual representation of the forged QR code does not match the network address of the malicious node, no traffic is visualized to or from the malicious node, which is noticed by the operator. Again, by observing the change in signal strength of legitimate packets while moving between the optically indistinguishable sensor nodes aids in identifying the malicious node.

Lastly, both visual and network representation of a sensor node can be forged. Network traffic originating from the forged node fails the signature check, resulting in no traffic being visualized at it. As before, EyeSec reports the failed signature verification attempts. Pointing the AR Device at the malicious node, EyeSec hints that this is the sensor node whose signature checks failed repeatedly.

To sum it up, MAC addresses are the common representation for networked devices, which can be used with least effort and maximum flexibility. However, cryptographic means must be utilized, otherwise attacks such as sensor node spoofing are possible. EyeSec has been designed such, that it can be used with both secured and insecure WSNs. It can handle cryptographic sensor node identities and aid the operator in detecting attacks resulting from spoofed sensor nodes.

EyeSec utilizes cryptography to prevent remote attacks over the network. The Backend as the central point of communication creates a wireless access point secured by Wi-Fi Protected Access Version 2 (WPA-2) and Pre-Shared Keys. All RESTful HTTP requests to the Backend require an Authentication Header. The authentication is done via user name and password. Every user name is assigned a role in the database at the Backend. For every API endpoint and every HTTP method, we specify, which roles are allowed to access this endpoint. If multiple Sniffer Units or AR Devices are used, each of them has individual login credentials. If a single device is compromised, we can revoke the compromised devices permissions. Further, we can provide fine-grained access to information at the Backend. A senior operator could be permitted to add new nodes to the Backend, while a trainee might be given read access only.

Currently, EyeSec is tailored towards WSN monitoring and troubleshooting. However, having incorporated protection against node spoofing attacks and by utilizing secure communications, EyeSec is fit for usage in a hostile environment. In our ongoing research, we are going to enhance EyeSec by adding detectors for various attacks on the WSN, helping operators to spot more attacks at a glance.

## 3 Implementation

In this section, we describe the proof of concept implementation of EyeSec. Figure 3 shows hardware and software modules as a block diagram, while Figure 4 shows the hardware units we used in the evaluation.

Our implementation currently supports the *6LoWPAN* transmission protocol using the *Routing Protocol for Low power and Lossy Networks (RPL)*, which creates a dynamic mesh network [16]. As EyeSec has been designed to be retrofitted to already deployed WSNs, the sensor network is not part of EyeSec and thus not discussed in this section.

### 3.1 Sniffer Unit

Sniffer Units capture 6LoWPAN traffic using a *CC2650 Launchpad* by *Texas Instruments* as transceiver, which utilizes the *Sensniff-Peripheral* firmware shipped with the Contiki-NG operating system [2]. Captured packets are transferred to a Raspberry Pi Zero W using UART at 460800 Baud. There, packets flow downwards the processing blocks displayed in Figure 3. First, the *Host-Sensniff* tool converts packets to PCAP format and forwards them to a pipe [6].

Next, to extract the digital device representation, we pass packets from the pipe into *pyshark*, a Python wrapper for *Tshark* [5]. There, the captured packets are being dissected into their layers. Lastly, a custom packet handler derives MAC addresses from each IPv6 address encountered in the captured packet. It extracts multiple hops and treats them as separate packets.

### 3.2 Backend

The Backend is implemented on a Raspberry Pi 3. It is shown in the middle of Figure 3. It creates a Wi-Fi access point and exposes its services, e.g. the database and time synchronization, over a web server, which is implemented with the *Flask* framework [24]. The Sniffer Unit and the AR Device can interact with the Backend using a RESTful API. For information storage, a *PostgreSQL* database managed by the *SQLAlchemy* toolkit is utilized [7, 4].

### 3.3 AR Device

A *Samsung Galaxy S7* smart phone running *Android OS* 8.0 is used as the handheld AR Device[1]. The smart phone runs the EyeSec App. In order to be easily extensible, the App is divided into three submodules: *Recognition*, *Core* and *Application*.

The camera of the smart phone captures the environment and shows the live image at its screen. For interacting with the camera, the *camera2* API is employed [1]. To detect and read QR codes, the *zxing* library is used [10]. Using the built-in camera of the smart phone, the 2 cm by 2 cm sized QR codes can be detected reliably at distances up to 1 m and angles up to $45°$. All functions concerning optical object recognition are grouped into the *Recognition* module. Thus, if an even more suitable method for visual device representation extraction will be available in the future, this module can be exchanged easily.

To interact with the Backend via the RESTful API, the *volley* framework is utilized [8]. Functions used for web service access are encapsulated in the module *Application*.

---
[1]EyeSec requires at least Android 6.0 to be able to leverage newly added cryptographic features.

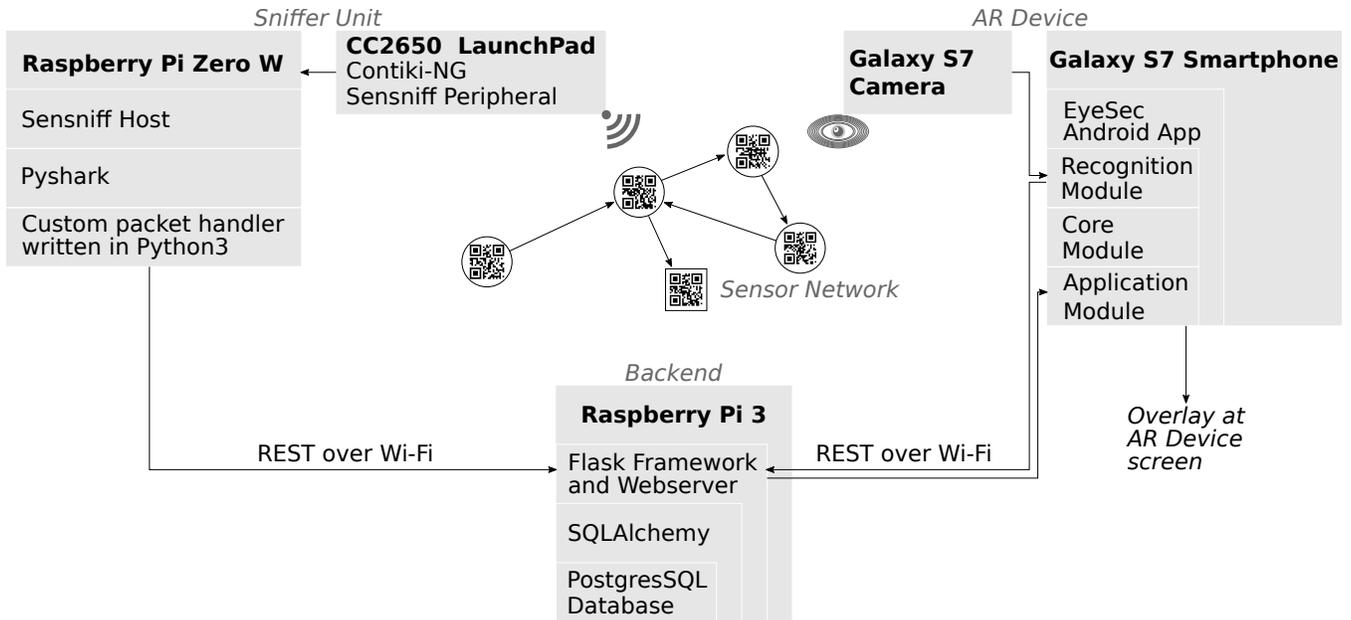

**Figure 3. Hardware and software building blocks and data flows between these.**

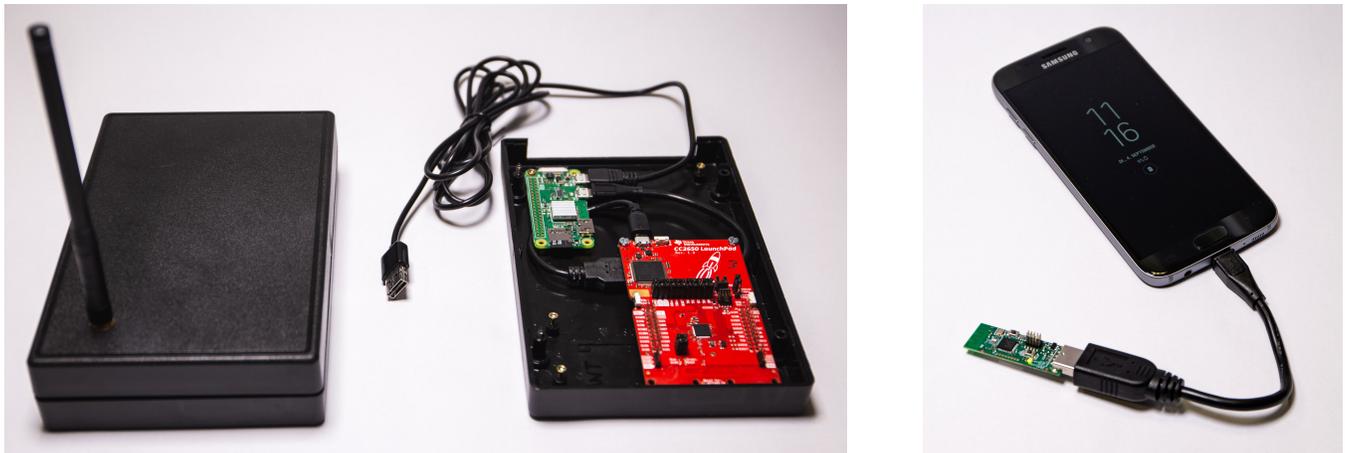

**Figure 4. Sniffer Units and a combined Sniffer Unit/AR Device used in the evaluation.**

If a sensor node has been identified by the camera, but was not found in the database at the Backend, the operator creates a new digital representation of that sensor node by using the AR Devices touch screen. Similarly, information on a existing sensor node (e.g. location) can be altered.

For creating the overlay on its screen, the AR Device requests data from the Backend. Our custom visualization uses Android graphics canvas objects to draw a line between different node views. The position of start and end of the line only depend on the current position of the node views. Hence, only little computational effort is needed to draw these views. Together with efficient QR code detection, AR Device battery is preserved. In order to visualize *asymmetric* connections, we draw a bezier curve between the two sensor nodes. An arrow indicates the direction of packet flow. Arrow stroke width increases in a logarithmic fashion with traffic density. Classes needed to draw the overlay and visualize the network data are encapsulated in the *Application* module, permitting easy modifications and replacements.

Figure 5 is a capture of the screen of the AR Device, showing the overlay. There are four sensor nodes (Sensortag 65, Launchpad 2, Sensortag 62 and Sensortag 63) and the WSN server (Launchpad 4). Each device is identified by its QR code and an overlay is placed above the QR code, showing the name of the device and additional information. Network traffic between sensor nodes and the server is shown as arrows. It can be seen in Figure 5 that there is bi-directional traffic, i.e. the server responds to a received packet.

## 4 Evaluation

In this section, we describe the usage of EyeSec in a real WSN. Our testbed consists of six sensor nodes using *CC2650 Sensortag* and *CC2650 Launchpad* microcontrollers. The sensor nodes transfer messages using the transmission protocol 6LoWPAN. Routing in the WSN is done by the *Routing Protocol for Low power and Lossy Networks (RPL)*, which creates a dynamic mesh network [16]. Sensor node firmware utilizes the *rpl-udp* example, which is part of the *Contiki-NG* operating system [2]. The example contains both a *client* and a *server* implementation. One CC2650 device is flashed with the server firmware, all others with client firmware.

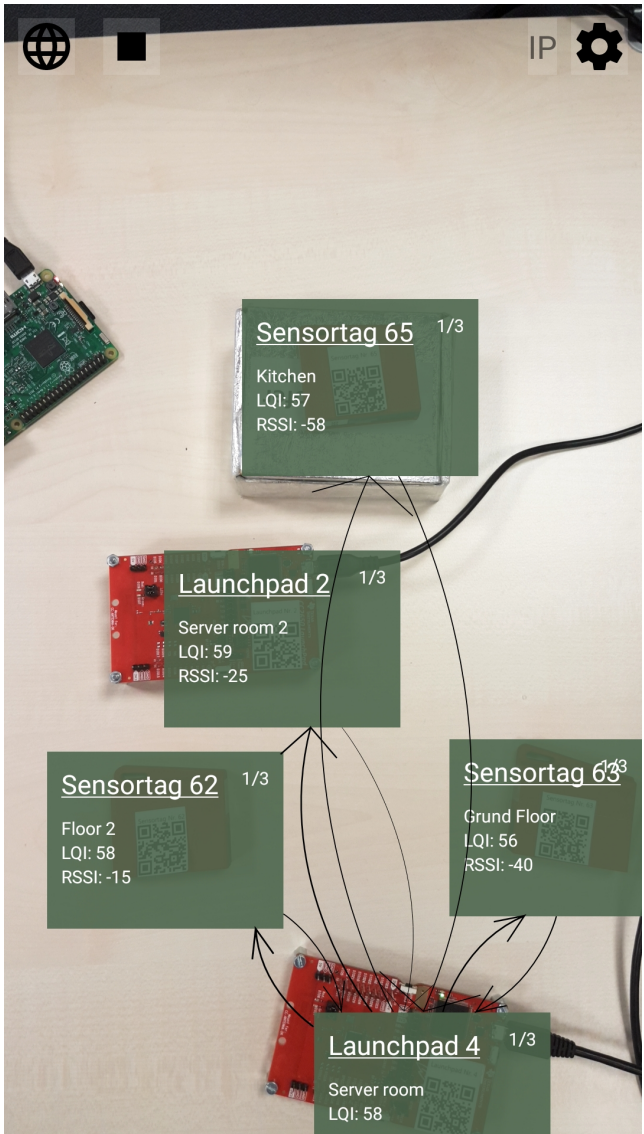 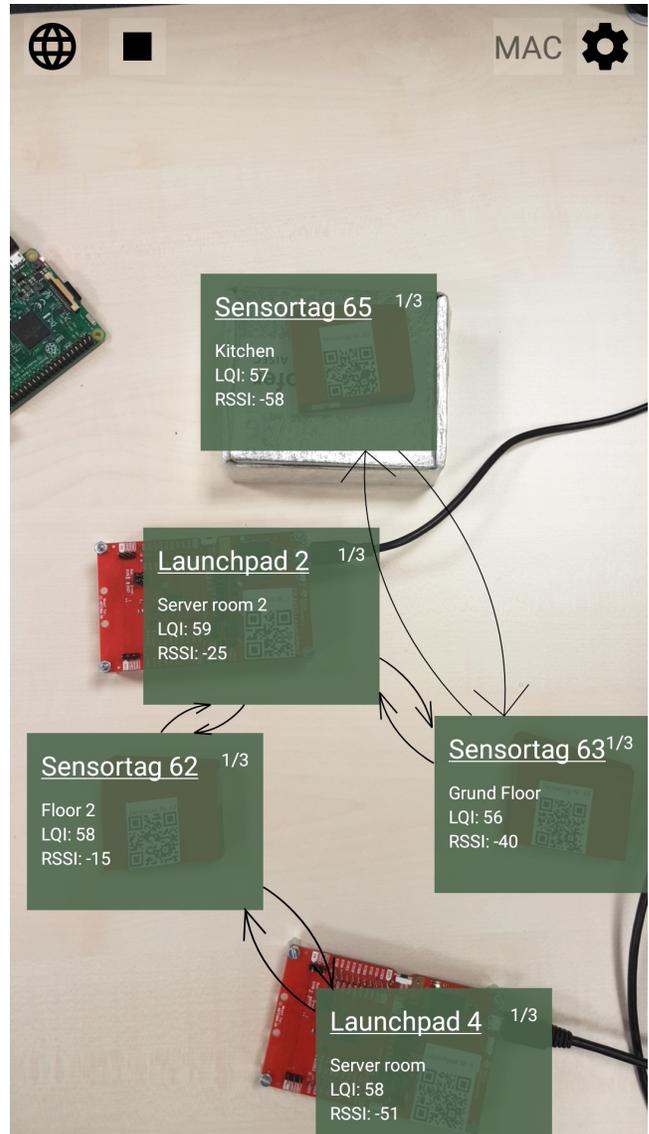

**Figure 5. Arrows indicate traffic between sensor nodes and WSN server (Launchpad 4).**

**Figure 6. MAC-address based traffic view reveals hops.**

Each client sends UDP packets with an increasing number in fixed transmit intervals of 10 sec to the server. The server replies with the same number to the client. CC2650 devices have been placed in an area of $1\,m^2$ inside an office building. Each CC2650 is supplied with a printed QR code containing the wireless interface MAC address of the device. No further modifications to the sensor network were performed.

To provide maximum mobility to the operator, the Sniffer Unit is combined with the AR Device into a single portable hand-held device. A separate Backend is placed such that Wi-Fi connectivity between Sniffer Unit/AR Device and Backend is given at all times.

### 4.1 Investigating Hop Behavior

While monitoring a 6LoWPAN network, EyeSec can show traffic either based on IP addressing or MAC addressing. The former is shown in Figure 5 and can be used to confirm that every sensor node has a connection to the WSN server (Launchpad 4). To gain deeper insight into the WSN, the operator can switch to MAC addressing to reveal the hops a packet takes. This is shown in Figure 6. It can be seen, that packets from device Sensortag 65 are forwarded by every other sensor node, until they reach their destination at Launchpad 4. If no arrows originate from or terminate at a sensor node, this indicates sensor node failure. Routing anomalies can be grasped intuitively, if e.g. some nodes exchange traffic but have no route to the server.

Besides helping the operator to identify sources of failure, the MAC-based view can be used to optimize network topology. The operator might identify a single sensor node, which forwards traffic from many sensor nodes towards the server. Such forwarder nodes are a potential source of network failure, as forwarding dense traffic drains their battery quickly. This brings the risk of network parts becoming isolated. To prevent this, the operator can e.g. re-arrange node positioning or add a new node to provide a second route to the server. Effects of those placement optimizations can be observed in real-time with EyeSec.

The operator can choose, how long arrows between communicating sensor nodes are displayed. With a short duration, single message transfers can be visualized. After the message transfer has finished, arrows disappear. Choosing a longer duration, traffic density can be visualized. Message transfers on a single route are summed up, increasing arrow thickness as more traffic is aggregated over time. This can be exploited to identify sensor nodes, whose traffic rates are unexpectedly high or low, which could indicate malfunction or an attack.

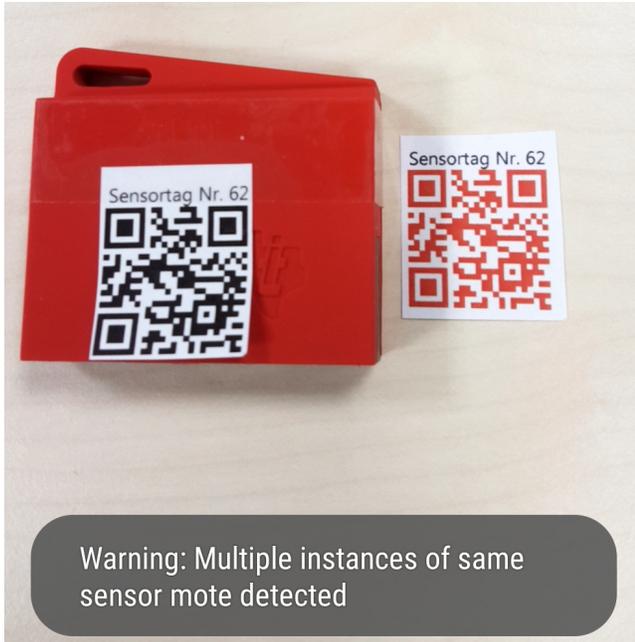

Figure 7. Duplicate QR Codes raise a warning.

### 4.2 Duplicate QR Codes

Duplicate QR codes can occur due to errors during creation and placement of those. Further, it can not be ruled out that there *are* duplicate MAC addresses in networked devices. This will eventually lead to collisions while routing network traffic. Duplicate QR codes can also be caused by an attack, in which an attacker tries to spoof the visual representation of another sensor node by intentionally placing duplicate QR codes. This could trick EyeSec into attributing traffic falsely. Thus, on detecting duplicate QR codes, EyeSec issues a warning and prevents the user from adding information to either sensor node identified as duplicate, until the ambiguity has been resolved. This warning is shown in Figure 7. It must be noted that in order to verify, which of two sensor nodes with the same MAC inscribed in the QR code is authentic, cryptographic solutions are needed. This has been discussed in Section 2.7.

### 4.3 Handling Sensor Nodes Out of Sight

Depending on sensor node placement, it can occur that only a single sensor node can be acquired by the AR Devices camera. To be still able to see traffic flows between the node in view and his neighbors, EyeSec tracks the AR Devices rotations and remembers, in which direction an adjacent sensor node has been seen previously. Thus, traffic flow can still be drawn between the sensor node, which is captured by the camera right now, and the last known position of another sensor node, not visible any more. This behavior is shown in Figure 8, where the AR Device has been rotated such that the neighbor of Sensortag 62 moved outside the lower boundary of the display. Arrows indicating message flow can still be seen between Sensortag 62 and the last known location of its neighbor. This behavior enables the operator to use EyeSec even in spatially extended WSNs, tracing hops in a 'bread-crumb' manner.

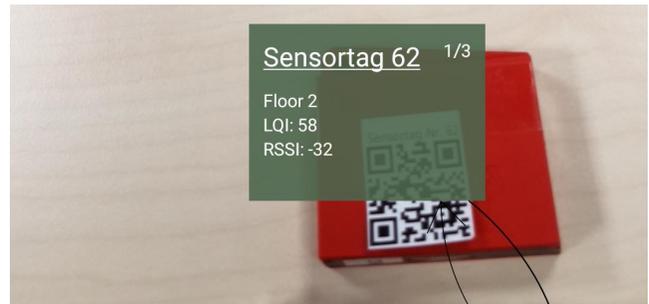

Figure 8. Traffic flows to devices not captured by the camera can still be shown.

### 4.4 Investigating Network Behavior over Time

As described in Section 2.1, captured packets are attributed with a time stamp. To be able to see how network topology has changed over time, the operator can utilize the full network graph view. An example is shown in Figure 9. This view gives an overview of the full network, showing all traffic captured either based on IP addressing as shown in Figure 9, or using MAC-based addressing, again revealing hops. The operator can step through time and e.g. pinpoint, when a particular sensor node has failed and which effects the failure had on network topology. By using this view together with the MAC-based overlay shown in Figure 6, a WSN problem can be traced down both temporally and spatially, supporting the troubleshooting process.

## 5 Related Work

EyeSec incorporates technologies from many domains, such as network data acquisition and information extraction, physical device detection and visualization techniques. We focus our discussion on works, which are concerned with passive network data acquisition and such, which focus on Augmented Reality-based visualization of networks.

Ringwald et al. discuss the design of a sniffer network for passive WSN observation [23, 22]. Unlike EyeSec, which lets the operator interpret what they see, Ringwald et al. try to analyze the cause for node failure. Further, their approach displays network topology and additional information on a central screen, being more similar to established network monitoring solutions. In contrast, EyeSec is tailored towards on-site WSN monitoring and troubleshooting.

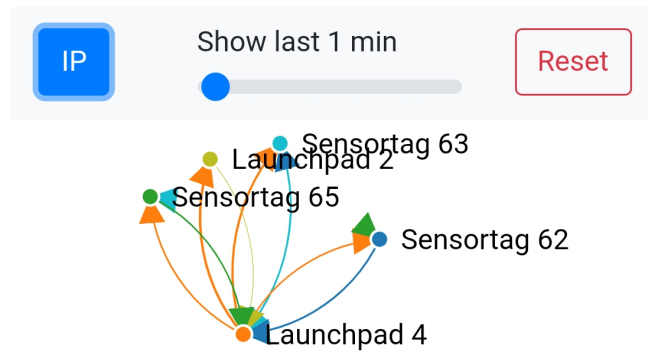

Figure 9. Full network graph permits viewing network behavior over time.

Rauhala et al. implement an AR interface for sensor networks on a hand-held device [21]. Their system visualizes readings of humidity sensors placed inside walls, the location of sensors being indicated by a unique marker placed at the wall. A similar system has been described by Goldsmith et al. [14]. Both approaches have in common, that the hand-held device, which visualizes sensor data, has been integrated into the sensor network system. Thus, retrofitting is not possible without altering the existing WSNs.

Sato et al. as well as Sakamoto et al. [27, 28, 26] present an AR-based approach, which visualizes links and traffic between devices as well as information on sensor nodes. They use a web cam for capturing the physical world and a touch screen for displaying the overlay. The touch screen permits interaction with the network. For example, the user is able to "cut" the connection between two devices by performing a cutting gesture through the virtual cable drawn in between. However, this interactivity comes at the cost of the AR hardware having to actively interact with the sensor network. Thus, the system can not be retrofitted to existing WSNs without modifying sensor node firmware.

Sahin et al. use an AR smart phone app to visualize antenna radiation patterns [25]. Devices are identified by optical markers. Their experimental study shows, that the marker-based device identification is a suitable approach and that visualization truly helps users to understand, what is happening. However, as they are concerned with radiation patterns rather than network monitoring, they require a specialized setup using Software-Defined Radios (SDR) wired to a processing server. Additionally, they need special software on the SDR to provide information on radiation patterns.

Ohta et al.'s approach places an overlay over sensor nodes, showing network topology and data flow [20]. They use marker-based node identification. Usability of their approach is limited by the fact that they require a wired connection between every mobile device and the visualization device. Additionally, software at each mobile device must be modified to provide information on the packets being sent. Lastly, each sensor node must be able to communicate with the visualization device at all times. This is a major drawback, since sensor nodes can fail for many reasons and the exact purpose of an AR visualization approach is to help operators identify and fix failed sensor nodes.

Koutitas et al. use a Microsoft HoloLens to show network topology visualizations [18, 17]. As they have only released a demo abstract, a set of slides and some demonstration videos, it is difficult to assess their system in detail. It seems, however that their approach is tailored towards a given ZigBee network. Additionally, they require dedicated AR headsets while EyeSec runs on commodity smart phones.

Compared to related works, EyeSec has been designed for usability and security. Unlike prior work, in which sensor nodes are queried actively, our contribution uses passive observation solely. This approach is significantly more challenging, as EyeSec can only extract relevant information from observed network traffic. Thus, our system has neither internal knowledge of the WSN, nor the possibility to directly query a sensor node for information. On the other hand, passive observation is beneficial, as it permits retrofitting our system. We consider this an important contribution for several reasons. Firstly, sensor node firmware does not need to be extended. Keeping firmware sleek, potential sources of error are reduced. This increases sensor node availability, which we consider an important property of WSNs. Secondly, by having less interface functions in firmware, potential entry points for attacks are minimized. This goes hand in hand with our detailed security concept, which to the best of our knowledge makes our system the first AR-based WSN monitoring tool designed with security in mind. Its modular low-cost hardware and software design utilizing Augmented Reality supplies the operator in the field with just the information needed to solve a specific problem.

## 6 Conclusion

We have presented the design and implementation of EyeSec, a protocol-agnostic and portable AR tool to aid network operators in analyzing and troubleshooting wireless sensor networks on-site. EyeSec utilizes one or more Sniffer Units, a Backend and one or more AR Devices. The mobile Sniffer Units are placed temporarily among sensor nodes. They passively capture network traffic, thus no modifications to a given WSN are needed. The low-cost and modular design of the Sniffer Units permits efficient coverage of spatially extended WSNs. Support for additional WSN transmission protocols can be added easily. Visual device representations are extracted from QR codes by the AR Device, implemented on a smart phone. The central Backend consolidates information extracted from network traffic and visual device information. Those are fed into the AR Device, which provides real time visualization of connections, traffic flows and multi-hop behavior.

We implemented a real sensor network utilizing 6LoWPAN transmission protocol. Using this network, we explored scenarios commonly encountered in WSN troubleshooting. Our results show, how EyeSec aids the operator in the field to pinpoint and fix those failures.

EyeSecs distinguishing feature is its high usability. To operators, WSN reliability is paramount. Thus, already deployed WSN firmware will not be modified just to support a new monitoring and troubleshooting tool, as all modifications can introduce additional sources of failure. Due to being retrofittable by operating passively and independently of existing infrastructure, wide acceptance can be expected. This enables EyeSec to be used in real WSNs, overcoming objections of WSN owners and field operators.

In future research, we are going to evaluate EyeSec in a large-scale WSN deployed in a factory environment. During this assessment, we are going to conduct interviews with operators to further optimize EyeSec towards their needs.

Currently, EyeSec can only handle valid packets. As strong interference in the factory environment can be expected, we are going to add support for malformed packets by processing raw bits received. Additionally, we are going to extend EyeSec for security research in WSN.